\documentstyle[12pt]{article}

\global\arraycolsep=2pt
\begin{document}

\begin{titlepage}

\begin{flushright}
SLAC--PUB--6258\\
July 1993\\
T/E
\end{flushright}

\vspace{0.3cm}

\begin{center}
\Large\bf Heavy Hadron Weak Decay Form Factors\\
to Next-to-Leading Order in 1/m$_{\bf Q}$
\end{center}

\vspace{0.8cm}

\begin{center}
Matthias Neubert\footnote{Supported by the Department of Energy
under contract DE-AC03-76SF00515.}\\
Stanford Linear Accelerator Center\\
Stanford University, Stanford, California 94309
\end{center}

\vspace{0.8cm}

\begin{abstract}
Based on the short-distance expansion of currents in the heavy quark
effective theory, we derive the exact expressions for the
heavy-to-heavy meson and baryon weak decay form factors to order
$1/m_Q$ in the heavy quark expansion, and to all orders in perturbation
theory. We emphasize that the Wilson coefficients in this expansion
depend on a kinematic variable $\bar w$ that is different from the
velocity transfer $w=v\cdot v'$ of the hadrons. Our results generalize
existing ones obtained in the leading-logarithmic approximation. Some
phenomenological applications are briefly discussed.
\end{abstract}

\centerline{(submitted to Nuclear Physics B)}

\end{titlepage}

\section{Introduction}

The heavy quark effective theory (HQET) is by now a well-established
tool to investigate the properties of hadrons containing a single heavy
quark \cite{Eich,Geor,Mann,FGL,Falk,habil}. Most important are its
applications to the weak decay form factors describing the semileptonic
transitions of the type $H_b\to H_c\,\ell\,\bar\nu$, from a study of
which one can extract the element $V_{cb}$ of the
Cabibbo-Kobayashi-Maskawa matrix \cite{Volo,Vcb}. Here $H_Q$ denotes a
hadron containing a single heavy quark $Q$. Based on a short-distance
expansion of the flavor-changing weak currents, HQET provides a
systematic expansion of the decay amplitudes in powers of $1/m_c$ and
$1/m_b$. At leading order the effective current operators are of the
form $\bar h_{v'}^c\,\Gamma\,h_v^b$, where $\Gamma$ denotes some Dirac
matrix with the same quantum numbers as the current in the full theory,
and $h_v^Q$ are velocity-dependent fields that represent the heavy
quarks in the effective theory. Isgur and Wise have shown that hadronic
matrix elements of such operators can be parameterized by universal
form factors that are independent of the flavor and spin of the heavy
quarks and of the Dirac structure of $\Gamma$ \cite{Isgu}. These
so-called Isgur-Wise functions only depend on the quantum numbers of
the cloud of light degrees of freedom surrounding the heavy quarks. In
particular, transitions between two ground-state heavy mesons
(pseudoscalar or vector) or baryons are each described by a single
Isgur-Wise function, which is usually denoted by $\xi(v\cdot v')$ is
the case of mesons, and $\zeta(v\cdot v')$ in the case of baryons. Here
$v$ and $v'$ are the hadron velocities, which in the $m_Q\to\infty$
limit coincide with the velocities of the heavy quarks. The physical
origin of this remarkable reduction of form factors is a spin-flavor
symmetry of the leading-order effective Lagrangian of HQET.

At order $1/m_Q$ (we use $m_Q$ as a generic notation for $m_c$ or
$m_b$), a complete set of dimension-four operators appears in the
short-distance expansion of the weak currents and of the effective
Lagrangian of HQET, which induce symmetry-breaking corrections to the
heavy quark limit. Additional form factors are required to parameterize
the matrix elements of these operators. Luke has shown that for meson
decays one needs to introduce four additional functions as well as a
mass parameter $\bar\Lambda$ \cite{Luke}, which can be interpreted as
the effective mass of the light degrees of freedom \cite{AMM}. In the
case of baryon transitions, the introduction of an analogous parameter
and a single new function are sufficient \cite{GGW}. However, although
the results of these papers are general and do in principle allow one
to obtain the exact expressions for weak decay form factors to order
$1/m_Q$, so far these expressions have not yet been derived. The reason
is that at order $1/m_Q$ the short-distance expansion of the currents
was not known beyond the leading-logarithmic approximation, which when
applied to the scaling between $m_b$ and $m_c$ is known to be a very
crude approximation of at best pedagogical relevance \cite{QCD}. To get
reliable numerical results, a full next-to-leading order calculation is
unavoidable. Such a calculation is already very tedious at leading
order in the heavy quark expansion, however, and to go to order $1/m_Q$
seemed very complicated.

In a recent paper we have shown that a ``hidden'' symmetry of HQET,
namely its invariance under reparameterizations of the heavy quark
momentum operator \cite{LuMa}, leads to the surprising result that all
the Wilson coefficients appearing at order $1/m_Q$ in the
short-distance expansion of the weak currents can be related to the
coefficients appearing at leading order \cite{RPI}. By virtue of this
result one can construct the expansion of any current to
next-to-leading order in $1/m_Q$, without even knowing the explicit
structure of the Wilson coefficients. For instance, the most general
form of the vector current is
\begin{eqnarray}\label{expand}
   \bar c\,\gamma^\alpha b &\to& C_1(\bar w,\mu)\,\bar h_{v'}^c\,
    \bigg[ \gamma^\alpha + \gamma^\alpha\,{i\,\rlap/\!D\over 2 m_b}
    - {i\overleftarrow{\rlap/\!D}\over 2 m_c}\,\gamma^\alpha
    \bigg]\,h_v^b \nonumber\\
   &+& C_2(\bar w,\mu)\,\bar h_{v'}^c\,\bigg[ v^\alpha
    + {i D^\alpha\over m_b} + v^\alpha\,{i\,\rlap/\!D\over 2 m_b}
    - {i\overleftarrow{\rlap/\!D}\over 2 m_c}\,v^\alpha
    \bigg]\,h_v^b \\
   &+& C_3(\bar w,\mu)\,\bar h_{v'}^c\,\bigg[ v'^\alpha
    - {i\overleftarrow{D}\!\!\phantom{|}^\alpha\over m_c}
    + v'^\alpha\,{i\,\rlap/\!D\over 2 m_b}
    - {i\overleftarrow{\rlap/\!D}\over 2 m_c}\,v'^\alpha
    \bigg]\,h_v^b + {\cal{O}}(1/m_Q^2) \,. \nonumber
\end{eqnarray}
A similar expansion with coefficients $C_i^5(\bar w,\mu)$, and with
$\gamma_5$ inserted after whatever object carries the Lorentz index
$\alpha$, obtains for the axial vector current. In these expressions,
the kinematic variable $\bar w$ to be used in the Wilson coefficients
is different from the velocity transfer $w=v\cdot v'$ of the hadrons.
The relation is \cite{RPI}
\begin{equation}\label{wbar}
   \bar w = w + \bigg( {\bar\Lambda\over m_b}
   + {\bar\Lambda\over m_c} \bigg) (w-1) + {\cal{O}}(1/m_Q^2) \,,
\end{equation}
where $\bar\Lambda$ is defined as the asymptotic mass
difference between a hadron $H_Q$ and the heavy quark that it contains.
In the $m_Q\to\infty$ limit, this difference approaches a finite value,
$m_{H_Q}-m_Q\to\bar\Lambda$, which can be identified with effective
mass of the light degrees of freedom in the hadronic bound
state.\footnote{This definition of $\bar\Lambda$ implicitly relies on a
particular choice of the heavy quark mass, as discussed in detail in
Ref.~\cite{AMM}.}

The variable $\bar w$ can be interpreted as the velocity transfer of
free quarks. Consider the weak transition $H_b\to H_c+W^-$. The bottom
quark in the initial state $H_b$ moves on average (up to fluctuations
of order $1/m_b$) with the hadron's velocity. When the $W$-boson is
emitted, the outgoing charm quark has in general some different
velocity. Let us denote the product of these velocities by $\bar w$.
Since over short time scales the quark velocities remain unchanged,
this is what is ``seen'' by hard gluons. After the $W$-emission, the
light degrees of freedom still have the initial hadron's velocity. But
they have to recombine with the outgoing heavy quark to form the final
state $H_c$. Thus, a rearrangement is necessary, which happens over
much larger, hadronic time scales by the exchange of soft gluons. In
this process the velocity of the charm quark is changed by an amount of
order $1/m_c$ (its momentum is changed by an amount of order
$\Lambda_{\rm QCD}$). Hence the final hadron velocity transfer $w$ will
differ from the ``short-distance'' quark velocity transfer $\bar w$ by
an amount of order $1/m_Q$. The precise relation between $w$ and $\bar
w$ is determined by momentum conservation and is given in (\ref{wbar}).
In fact, this relation is nothing but the condition
$(p_{H_b}-p_{H_c})^2 = (p_b-p_c)^2$, i.e., that the momentum transfer
to the hadrons equals the momentum transfer to the heavy quarks. At
zero recoil, no rearrangement is needed, and indeed $w=\bar w=1$ in
this limit.

In the case of flavor-changing currents, the Wilson coefficients in
(\ref{expand}) are very complicated functions of $\bar w$, the heavy
quark masses, and the renormalization scale $\mu$. For the moment, the
precise structure of these functions is not important. What will be
relevant is that the $\mu$-dependence can be factorized into a
universal function $K_{\rm hh}(\bar w,\mu)$, which is independent of
the heavy quark masses and is the same for any bilinear heavy quark
current in HQET:
\begin{equation}\label{fact}
   C_i^{(5)}(\bar w,\mu) = \widehat{C}_i^{(5)}(m_b,m_c,\bar w)\,
   K_{\rm hh}(\bar w,\mu) \,.
\end{equation}
At zero recoil this function is independent of $\mu$, and it is
convenient to use a normalization such that $K_{\rm hh}(1,\mu)=1$. From
now on we shall for simplicity write $\widehat{C}_i^{(5)}(\bar w)$ and
not display the dependence on the heavy quark masses.

In addition to the currents, one needs the effective Lagrangian of HQET
to order $1/m_Q$. For a heavy quark $Q$ with velocity $v$, it reads
\cite{Eich,Geor,Mann,FGL}
\begin{equation}\label{Leff}
   {\cal{L}}_{\rm eff} = \bar h_v^Q\,i v\cdot D\,h_v^Q
   + {1\over 2 m_Q}\,{\cal{L}}_1^Q + {\cal{O}}(1/m_Q^2) \,,
\end{equation}
where
\begin{equation}
   {\cal{L}}_1^Q = \bar h_v^Q\,(i D)^2\,h_v^Q
   + C_{\rm mag}(\mu)\,{g_s\over 2}\,\bar h_v^Q\,
   \sigma_{\alpha\beta}\,G^{\alpha\beta}\,h_v^Q \,.
\end{equation}
Similar to (\ref{fact}), the Wilson coefficient of the chromo-magnetic
operator can be written in the factorized form
\begin{equation}
   C_{\rm mag}(\mu) = \widehat{C}_{\rm mag}(m_Q)\,K_{\rm mag}(\mu) \,,
\end{equation}
where $K_{\rm mag}(\mu)$ is independent of $m_Q$. The first operator in
${\cal{L}}_1^Q$ is not renormalized \cite{LuMa}.

Explicit expressions for the short-distance coefficients can be found
in the literature (see, e.g., Refs.~\cite{Eich,FGL,QCD}). With the
exception of $C_{\rm mag}$, they are known to next-to-leading order in
renormalization-group improved perturbation theory. However, as written
above the short-distance expansions (\ref{expand}) and (\ref{Leff}) are
completely general and true to all orders in perturbation theory. Given
these results, one can derive the exact expressions for the weak decay
form factors of heavy mesons or baryons to order $1/m_Q$. This is the
purpose of this paper. The resulting expressions are presented in an
explicitly renormalization-group invariant form by introducing
$\mu$-independent Isgur-Wise functions and Wilson coefficients. They
generalize approximate results obtained in leading-logarithmic
approximation by Luke \cite{Luke} and Georgi et al.~\cite{GGW}. In
Sec.~\ref{sec:2} we discuss the form factors for the decay
$\Lambda_b\to\Lambda_c\,\ell\, \bar\nu$. The more complicated case of
meson weak decays is considered in Sec.~\ref{sec:3}.

\section{Baryon Form Factors}
\label{sec:2}

Consisting of a heavy quark and light degrees of freedom with quantum
numbers of a spin-0 diquark, the ground-state $\Lambda_Q$ baryons are
particularly simple hadrons. In HQET they are represented by a spinor
$u_\Lambda(v,s)$ that can be identified with the spinor of the heavy
quark. For simplicity, we shall from now on omit the spin labels and
write $u(v)\equiv u_\Lambda(v,s)$. The baryon matrix elements of the
weak currents $V^\alpha=\bar c\,\gamma^\alpha b$ and $A^\alpha=\bar c\,
\gamma^\alpha \gamma_5\,b$ can be parameterized by six hadronic form
factors, which we write as functions of the baryon velocity transfer
$w=v\cdot v'$:
\begin{eqnarray}
   \langle \Lambda_c(v')|\,V^\alpha\,|\Lambda_b(v)\rangle
   &=& \bar u(v') \Big[ F_1(w)\,\gamma^\alpha + F_2(w)\,v^\alpha
    + F_3(w)\,v'^\alpha \Big]\,u(v) \,, \nonumber\\
   \phantom{ \bigg[ }
   \langle \Lambda_c(v')|\,A^\alpha\,|\Lambda_b(v)\rangle
   &=& \bar u(v') \Big[ G_1(w)\,\gamma^\alpha + G_2(w)\,v^\alpha
    + G_3(w)\,v'^\alpha \Big] \gamma_5\,u(v) \,. \nonumber\\
   &&
\end{eqnarray}
The aim is to construct an expansion of $F_i(w)$ and $G_i(w)$ in powers
of $1/m_Q$, and to relate the coefficients in this expansion to
universal, mass-independent functions of the velocity transfer. Given
the operator product expansion of the weak currents as in
(\ref{expand}), this is achieved by evaluating the matrix elements of
the effective current operators. The baryon matrix elements of the
dimension-three operators can be parameterized in terms of a single
Isgur-Wise function $\zeta(w,\mu)$ defined by \cite{IW91,Ge91,MR91,Hu91}
\begin{equation}
   \langle\Lambda_c(v')|\,\bar h_{v'}^c\,\Gamma\,h_v^b\,
   |\Lambda_b(v)\rangle = \zeta(w,\mu)\,\bar u(v')\,\Gamma\,u(v) \,.
\end{equation}
As discussed by Georgi, Grinstein, and Wise \cite{GGW}, the power
corrections of order $1/m_Q$ involve contributions of two types. The
first come from the dimension-four operators in the expansion of the
currents. Their matrix elements can be related to the generic matrix
element
\begin{equation}
   \langle\Lambda_c(v')|\,\bar h_{v'}^c\,\Gamma\,i D_\beta\,h_v^b\,
   |\Lambda_b(v)\rangle = \zeta_\beta(v,v',\mu)\,
   \bar u(v')\,\Gamma\,u(v) \,.
\end{equation}
The equation of motion $i v\!\cdot\!D\,h_v^Q=0$ allows one to relate
$\zeta_\beta(v,v',\mu)$ to the leading-order Isgur-Wise function:
\begin{equation}
   \zeta_\beta(v,v',\mu) = {w\,v_\beta - v'_\beta\over w+1}\,
   \bar\Lambda\,\zeta(w,\mu) \,.
\end{equation}
The form factors also receive corrections from insertions of the
higher-di\-men\-sion operators of the effective Lagrangian into matrix
elements of the leading-order currents. However, baryon matrix elements
with an insertion of the chromo-magnetic operator vanish, since the
total spin of the light degrees of freedom is zero. Insertions of the
kinetic operator preserve the Dirac structure of the currents, hence
effectively correcting the Isgur-Wise function $\zeta(w,\mu)$. The
total effect is
\begin{equation}
   \langle\Lambda_c(v')|\,i\!\int\!{\rm d}x\,T\,\big\{\,
   \bar h_{v'}^c\,\Gamma\,h_v^b(0),{\cal{L}}_1^b(x) \,\big\}\,
   |\Lambda_b(v)\rangle = 2\bar\Lambda\,\chi(w,\mu)\,
   \bar u(v')\,\Gamma\,u(v) \,.
\end{equation}
We have factored out $\bar\Lambda$ to obtain a dimensionless form
factor $\chi(w,\mu)$. Insertions of ${\cal{L}}_1^c$ can be
parameterized by the same function.

Given these definitions, one can readily work out the explicit
expressions for the hadronic form factors $F_i(w)$ and $G_i(w)$ at
next-to-leading order in the $1/m_Q$ expansion. Of course, the physical
form factors should be written in terms of renormalized universal
functions rather than the $\mu$-dependent functions that parameterize
the matrix elements in the effective theory. According to (\ref{fact}),
the $\mu$-dependence of the Wilson coefficients of any bilinear heavy
quark current can be factorized into a universal function $K_{\rm
hh}(\bar w,\mu)$, which is normalized at zero recoil. It is now
important to recall that the variable $\bar w$ differs from the hadron
velocity transfer by terms of order $1/m_Q$. Using (\ref{wbar}) we find
\begin{equation}
   K_{\rm hh}(\bar w,\mu) = K_{\rm hh}(w,\mu)
   + \bigg({\bar\Lambda\over m_b} + {\bar\Lambda\over m_c}\bigg)\,
   (w-1)\,{\partial\over\partial w}\,K_{\rm hh}(w,\mu) + \ldots \,.
\end{equation}
The second term gives an extra contribution to the renormalization of
$\chi(w,\mu)$. We define
\begin{equation}
   K_{\rm hh}(\bar w,\mu)\,\bigg[ \zeta(w,\mu)
   + \bigg({\bar\Lambda\over m_b} + {\bar\Lambda\over m_c}\bigg)\,
   \chi(w,\mu) \bigg] \equiv \zeta_{\rm ren}(w)
   + \bigg({\bar\Lambda\over m_b} + {\bar\Lambda\over m_c}\bigg)\,
   \chi_{\rm ren}(w) \,,
\end{equation}
so that
\begin{eqnarray}\label{renfun}
   \zeta_{\rm ren}(w) &=& \zeta(w,\mu)\,K_{\rm hh}(w,\mu) \,,
    \nonumber\\
   \chi_{\rm ren}(w) &=& K_{\rm hh}(w,\mu)\,\chi(w,\mu)
   + \zeta_{\rm ren}(w)\,(w-1)\,{\partial\over\partial w}
   \ln K_{\rm hh}(w,\mu) \,.
\end{eqnarray}
Note that, because of $K_{\rm hh}(1,\mu)=1$, the renormalized universal
functions agree with the original functions at zero recoil.

For the presentation of our results we find it convenient to introduce
the dimensionless ratios $\varepsilon_Q = \bar\Lambda/2 m_Q$, and to
collect certain $1/m_Q$ corrections that always appear in combination
with the Isgur-Wise function into a new function
\begin{equation}
   \widehat{\zeta}_{bc}(w) \equiv \zeta_{\rm ren}(w) +
   (\varepsilon_c + \varepsilon_b)\,\bigg[ 2 \chi_{\rm ren}(w)
   + {w-1\over w+1}\,\zeta_{\rm ren}(w) \bigg] \,.
\end{equation}
Because of the dependence on the heavy quark masses this is no longer a
universal form factor. However, the flavor dependence is irrelevant as
long as one considers $\Lambda_b\to\Lambda_c$ transitions only.
Furthermore, we shall see below that $\widehat{\zeta}_{bc}(w)$ and
$\zeta_{\rm ren}(w)$ obey the same normalization at zero recoil. Let us
factorize the hadronic form factors according to
\begin{equation}
   F_i(w) = N_i(w)\,\widehat{\zeta}_{bc}(w) \,,\qquad
   G_i(w) = N_i^5(w)\,\widehat{\zeta}_{bc}(w) \,.
\end{equation}
Then the exact next-to-leading order expressions for the correction
factors are:
\begin{eqnarray}\label{Rires}
   N_1(w) &=& \widehat{C}_1(\bar w)\,\bigg[ 1 + {2\over w+1}\,
    (\varepsilon_c + \varepsilon_b) \bigg] \,, \nonumber\\
   N_2(w) &=& \widehat{C}_2(\bar w)\,\bigg( 1 +
    {2w\,\varepsilon_b\over w+1} \bigg)
    - \Big[ \widehat{C}_1(\bar w) + \widehat{C}_3(\bar w) \Big]\,
    {2\varepsilon_c\over w+1} \,, \nonumber\\
   N_3(w) &=& \widehat{C}_3(\bar w)\,\bigg( 1 +
    {2w\,\varepsilon_c\over w+1} \bigg)
    - \Big[ \widehat{C}_1(\bar w) + \widehat{C}_2(\bar w) \Big]\,
    {2\varepsilon_b\over w+1} \,, \nonumber\\
   && \\
   \phantom{ \bigg[ } N_1^5(w) &=& \widehat{C}_1^5(\bar w)
    \,, \nonumber\\
   N_2^5(w) &=& \widehat{C}_2^5(\bar w)\,\bigg( 1
    + {2\varepsilon_c\over w+1} + 2\varepsilon_b \bigg)
    - \Big[ \widehat{C}_1^5(\bar w) + \widehat{C}_3^5(\bar w) \Big]\,
    {2\varepsilon_c\over w+1} \,, \nonumber\\
   N_3(w) &=& \widehat{C}_3^5(\bar w)\,\bigg( 1 + 2\varepsilon_c
   + {2\varepsilon_b\over w+1} \bigg)
    + \Big[ \widehat{C}_1^5(\bar w) - \widehat{C}_2^5(\bar w) \Big]\,
    {2\varepsilon_b\over w+1} \,. \nonumber
\end{eqnarray}
It is remarkable that, up to an overall unknown function
$\widehat{\zeta}_{bc}(w)$, the baryon form factors at order $1/m_Q$ are
completely determined in terms of $\varepsilon_i$ and the
short-distance coefficient functions. Notice that the Wilson
coefficients are functions of the ``short-distance'' quark velocity
transfer $\bar w$, whereas the remaining kinematic expressions and the
universal form factors depend on the hadron velocity transfer $w$. In
leading-logarithmic approximation, where $\widehat{C}_1=
\widehat{C}_1^5$ and all other coefficients are set to zero, our exact
expressions reduce to the approximate results obtained in
Ref.~\cite{GGW}.

\begin{table}[htb]
   \vspace{0.5cm}
   \centerline{\begin{tabular}{c|ccc|ccc|ccc}
$\bar w$ & $w_{\Lambda_b\to\Lambda_c}$ & $w_{\bar B\to D^*}$ &
$w_{\bar B\to D}$ & $\widehat{C}_1$ & $\widehat{C}_2$ &
$\widehat{C}_3$ & $\widehat{C}_1^5$ & $\widehat{C}_2^5$ &
$\widehat{C}_3^5$ \\
\hline
1.0 & 1.00 & 1.00 & 1.00 & 1.14 & $-0.08$ & $-0.02$ & 0.99 &
 $-0.12$ & 0.04 \\
1.1 & 1.05 & 1.06 & 1.07 & 1.11 & $-0.08$ & $-0.02$ & 0.97 &
 $-0.11$ & 0.04 \\
1.2 & 1.11 & 1.13 & 1.15 & 1.08 & $-0.08$ & $-0.02$ & 0.95 &
 $-0.11$ & 0.04 \\
1.3 & 1.16 & 1.19 & 1.22 & 1.06 & $-0.07$ & $-0.02$ & 0.93 &
 $-0.10$ & 0.04 \\
1.4 & 1.22 & 1.25 & 1.29 & 1.03 & $-0.07$ & $-0.02$ & 0.91 &
 $-0.10$ & 0.03 \\
1.5 & 1.27 & 1.31 & 1.37 & 1.01 & $-0.07$ & $-0.02$ & 0.89 &
 $-0.09$ & 0.03 \\
1.6 & 1.33 & 1.38 & 1.44 & 0.99 & $-0.06$ & $-0.02$ & 0.88 &
 $-0.09$ & 0.03 \\
1.7 & 1.38 & 1.44 & 1.51 & 0.97 & $-0.06$ & $-0.02$ & 0.86 &
 $-0.09$ & 0.03 \\
1.8 & 1.43 & 1.50 & 1.59 & 0.95 & $-0.06$ & $-0.02$ & 0.85 &
 $-0.08$ & 0.03
   \end{tabular}}
   \centerline{\parbox{12cm}{\caption{\label{tab:1}
Short-distance coefficients for $b\to c$ transitions.
   }}}
\end{table}

For the numerical evaluation of the correction factors $N_i^{(5)}(w)$
we use the next-to-leading order expressions for the Wilson
coefficients from Ref.~\cite{QCD}. As input parameters we take
$m_b=4.80$ GeV and $m_c=1.45$ GeV for the heavy quark masses, and
$\Lambda_{\overline{\rm MS}}=0.25$ GeV (for $n_f=4$) in the two-loop
expressions for the running coupling constant. The resulting values of
the short-distance coefficients for different values of the quark
velocity transfer $\bar w$ are given in Table~\ref{tab:1}. For the
physical processes of interest, we also show the corresponding values
of the hadron velocity transfer $w$. They are obtained from
(\ref{wbar}) by using $\bar\Lambda_{\rm baryon}=0.84$ GeV and
$\bar\Lambda_{\rm meson}=0.51$ GeV, which give the correct hadron
masses. The corresponding values of the parameters $\varepsilon_Q$ for
$\Lambda_Q$ baryons are $\varepsilon_c\approx 0.29$ and
$\varepsilon_b\approx 0.09$. From (\ref{Rires}) we then obtain the
results shown in Table~\ref{tab:2}. The correction factors $N_i^{(5)}$
are given in dependence of the baryon velocity transfer $w$ over the
kinematic region accessible in $\Lambda_b\to\Lambda_c\,\ell\,\bar\nu$
decays. We find that symmetry-breaking corrections can be quite sizable
in heavy baryon decays. This is not too surprising, since
$\varepsilon_c\approx 0.3$ sets the natural scale of power corrections,
and the QCD corrections are typically of order
$\alpha_s(m_c)/\pi\approx 0.1$. We note, however, that upon contraction
with the lepton current the form factors $F_2(w)$ and $G_2(w)$ become
suppressed, relative to $F_3(w)$ and $G_3(w)$, by a factor
$m_{\Lambda_c}/m_{\Lambda_b}\approx 0.4$. The apparently large
corrections to these form factors thus become less important when one
computes physical decay amplitudes.

\begin{table}[htb]
   \vspace{0.5cm}
   \centerline{\begin{tabular}{c|ccc|ccc}
$w$ & $\sum_i N_i$ & $N_2$ & $N_3$ & $N_1^5$ & $N_2^5$ & $N_3^5$ \\
\hline
1.00 & 1.03 & $-0.42$ & $-0.12$ & 0.99 & $-0.48$ & 0.17 \\
1.11 & 0.98 & $-0.37$ & $-0.11$ & 0.94 & $-0.43$ & 0.15 \\
1.22 & 0.94 & $-0.34$ & $-0.10$ & 0.91 & $-0.39$ & 0.14 \\
1.33 & 0.90 & $-0.31$ & $-0.09$ & 0.88 & $-0.35$ & 0.13 \\
1.44 & 0.87 & $-0.29$ & $-0.09$ & 0.85 & $-0.32$ & 0.12
   \end{tabular}}
   \centerline{\parbox{12cm}{\caption{\label{tab:2}
Correction factors for the $\Lambda_b\to\Lambda_c$ decay form factors.
   }}}
\end{table}

Vector current conservation implies that $\sum_i F_i(1)=1$ in the limit
of equal baryon masses. Taking into account that $\sum_i
\widehat{C}_i(1)=1$ in this limit \cite{QCD}, we conclude that
$\zeta_{\rm ren}(1) + 4\varepsilon_Q\chi_{\rm ren}(1)=1$, which must be
satisfied for any value of $m_Q$. Hence, we recover the well-known
normalization conditions $\zeta_{\rm ren}(1)=1$ and $\chi_{\rm ren}(1)=
0$. It also follows that $\widehat{\zeta}_{bc}(1)=1$, which justifies
the definition of this function in the first place. At zero recoil, the
normalization of the baryon form factors is thus completely determined
by (\ref{Rires}). The most important consequence of these relations is
that the quantities $\sum_i F_i(w)$ and $G_1(w)$ do not receive any
$1/m_Q$ corrections at zero recoil. This is the analog of Luke's
theorem for baryon decays \cite{GGW}. Because of this result, it might
be possible to extract an accurate value of $V_{cb}$ from the
measurement of semileptonic $\Lambda_b$ decays near zero recoil, where
the decay rate is governed by the form factor $G_1$ alone. The
deviations from the prediction $G_1(1)=\widehat{C}_1^5(1)\approx 0.99$
are of order $1/m_Q^2$ and are expected to be small \cite{FaNe}.

\section{Meson Form Factors}
\label{sec:3}

The most important application of heavy quark symmetry is to derive
relations between the form factors parameterizing the exclusive weak
decays $\bar B\to D\,\ell\,\bar\nu$ and $\bar B\to D^*\ell\,\bar\nu$. A
detailed theoretical understanding of these processes is a necessary
prerequisite for a reliable determination of the element $V_{cb}$ of
the quark mixing matrix. We start by introducing a convenient set of
six hadronic form factors $h_i(w)$, which parameterize the relevant
meson matrix elements of the flavor-changing vector and axial vector
currents $V^\alpha=\bar c\,\gamma^\alpha b$ and $A^\alpha=\bar
c\,\gamma^\alpha\gamma_5\,b$:
\begin{eqnarray}\label{hdef}
   \phantom{ \Big[ }
   \langle D(v') |\,V^\alpha\,| \bar B(v) \rangle &=& h_+(w)\,(v+v')^\alpha
    + h_-(w)\,(v-v')^\alpha \,, \nonumber\\
   \phantom{ \bigg[ }
   \langle D^*(v',\epsilon) |\,V^\alpha\,| \bar B(v) \rangle &=&
    i\,h_V(w)\,\epsilon^{\alpha\beta\mu\nu}\,\epsilon_\beta^*\,
    v'_\mu\,v_\nu \,, \\
   \langle D^*(v',\epsilon) |\,A^\alpha\,| \bar B(v) \rangle &=&
    h_{A_1}(w)\,(w+1)\,\epsilon^{\ast\alpha} \!-\! \Big[
    h_{A_2}(w)\,v^\alpha + h_{A_3}(w)\,v'^\alpha \Big]
    \epsilon^*\!\cdot\!v \,. \nonumber
\end{eqnarray}
Here $w=v\cdot v'$ is the velocity transfer of the mesons. For
simplicity we work with a nonrelativistic normalization of states. To
obtain the standard relativistic normalization one has to multiply the
right-hand sides of (\ref{hdef}) by $\sqrt{m_B\, m_{D^{(*)}}}$.

In HQET, the doublet of the ground-state pseudoscalar and vector mesons
can be represented by a combined tensor wave function
\begin{equation}
   {\cal{M}}(v) = {1+\rlap/v\over 2}\,
   \cases{ -\gamma_5 \,; &pseudoscalar meson, \cr
           \rlap/\epsilon \,; &vector meson. \cr}
\end{equation}
We shall use a notation where $M(v)$ represents $\bar B$ or $\bar B^*$,
and $M'(v')$ stands for $D$ or $D^*$. Meson matrix elements of the
leading-order currents can then be written as \cite{Falk}
\begin{equation}
   \langle M'(v')|\,\bar h_{v'}^c\,\Gamma\,h_v^b\,|M(v)\rangle
   = - \xi(w,\mu)\,{\rm Tr}\Big\{\, \overline{\cal{M}}'(v')\,
   \Gamma\,{\cal{M}}(v) \Big\} \,,
\end{equation}
where $\xi(w,\mu)$ is {\sl the\/} Isgur-Wise function. The $1/m_Q$
corrections have been analyzed by Luke \cite{Luke}. Matrix elements of
the dimension-four current operators in (\ref{expand}) can be related to
\begin{equation}
   \langle M'(v')|\,\bar h_{v'}^c\,\Gamma\,i D_\beta\,h_v^b\,
   |M(v)\rangle = - {\rm Tr}\Big\{\, \xi_\beta(v,v',\mu)\,
   \overline{\cal{M}}'(v')\,\Gamma\,{\cal{M}}(v) \Big\} \,.
\end{equation}
The tensor form factor $\xi_\beta(v,v',\mu)$ has components
proportional to $v_\beta$, $v'_\beta$, and $\gamma_\beta$. The equation
of motion yields two relations among these three, and the final result
can be written in the form
\begin{equation}
   \xi_\beta(v,v',\mu) = {\bar\Lambda\over w+1}\,\xi(w,\mu)\,
   \bigg\{ \Big[ w - \eta(w) \Big]\,v_\beta - \Big[ 1 + \eta(w)
   \Big]\,v'_\beta - (w+1)\,\eta(w)\,\gamma_\beta \bigg\} \,,
\end{equation}
where $\eta(w)$ is a renormalization-group invariant function
\cite{xi3}.\footnote{In the notation of Ref.~\cite{Luke}, one has
$\bar\Lambda\,\eta(w)=\xi_3(w,\mu)/\xi(w,\mu)$.} A second class of
$1/m_Q$ corrections comes from insertions of higher-dimension operators
of the effective Lagrangian. The corresponding matrix elements have the
structure
\begin{eqnarray}\label{chidef}
   &&\langle M'(v')|\,i\!\int\!{\rm d}x\,T\,\big\{\,
    \bar h_{v'}^c\,\Gamma\,h_v^b(0),{\cal{L}}_1^b(x) \,\big\}\,
    | M(v)\rangle \\
   &&\qquad = - 2\bar\Lambda\,\chi_1(w,\mu)\,{\rm T}\Big\{\,
    \overline{\cal{M}}'(v')\,\Gamma\,{\cal{M}}(v) \,\Big\}
    \nonumber\\
   \phantom{ \int } &&\qquad\phantom{ = }
    - 2\bar\Lambda\,C_{\rm mag}(\mu)\,{\rm Tr}\Big\{\,
    \chi_{\alpha\beta}(v,v',\mu)\,\overline{\cal{M}}'(v')\,\Gamma\,
    P_+\,\sigma^{\alpha\beta} {\cal{M}}(v) \,\Big\} \,, \nonumber
\end{eqnarray}
and similar for an insertion of ${\cal{L}}_1^c$. Here
$P_+=\frac{1}{2}(1+\rlap/v)$. Again we have factored out $\bar\Lambda$
in order for the form factors to be dimensionless. The kinetic operator
contained in ${\cal{L}}_1^b$ transforms as a Lorentz scalar. An
insertion of it does not affect the Dirac structure of the matrix
element. Hence, the corresponding function $\chi_1(w,\mu)$ effectively
corrects the Isgur-Wise function. The chromo-magnetic operator, on the
other hand, carries a nontrivial Dirac structure. An insertion of it
brings a matrix $\sigma^{\alpha\beta}$ next to the meson wave function
${\cal{M}}(v)$. In addition, a propagator separates this insertion from
the heavy quark current, resulting in a projection operator $P_+$ to
the right of $\Gamma$. This explains the structure of the second trace
in (\ref{chidef}). Because of $v_\alpha\,P_+\,\sigma^{\alpha\beta}
{\cal{M}}(v)=0$, the most general decomposition of the tensor form
factor $\chi_{\alpha\beta}(v,v',\mu)$ is
\begin{equation}
   \chi_{\alpha\beta}(v,v',\mu) = i\,\chi_2(w,\mu)\,
   v'_\alpha\gamma_\beta + \chi_3(w,\mu)\,\sigma_{\alpha\beta} \,.
\end{equation}
The terms proportional to $\chi_3(w,\mu)$ in (\ref{chidef}) can be
simplified by means of the identity $P_+\,\sigma^{\alpha\beta}
{\cal{M}}(v)\,\sigma_{\alpha\beta} = 2 d_M\,{\cal{M}}(v)$, where
$d_P=3$ for pseudoscalar and $d_V=-1$ for vector mesons. It follows
that, irrespective of the structure of the current, the function
$\chi_3(w,\mu)$ always appears in combination with the Isgur-Wise
function, but with a coefficient that is different for pseudoscalar and
vector mesons. It thus represents a spin-symmetry violating correction
to the meson wave function.

What remains to be done is to introduce renormalized form factors. We
define $\xi_{\rm ren}(w)$ and $\chi_1^{\rm ren}(w)$ in analogy to
$\zeta_{\rm ren}(w)$ and $\chi_{\rm ren}(w)$ in (\ref{renfun}). For the
remaining functions, we define
\begin{equation}
   \chi_i^{\rm ren}(w) = K_{\rm mag}(\mu)\,K_{\rm hh}(w,\mu)\,
   \chi_i(w,\mu) \,;\qquad i=2,3.
\end{equation}
It is again convenient to introduce quantities $N_i(w)$, which contain
the sym\-me\-try-breaking corrections to the heavy quark limit, by
\begin{equation}
   h_i(w) = N_i(w)\,\xi_{\rm ren}(w) \,.
\end{equation}
As in the baryon case we denote $\varepsilon_Q = {\bar\Lambda/2 m_Q}$.
We stress, though, that the numerical values of these parameters are
different in the two cases. It is useful to define three new functions
$L_{P,V}(w)$ and $L_3(w)$ by
\begin{eqnarray}
   \xi_{\rm ren}(w)\,L_P(w) &=& 2 \chi_1^{\rm ren}(w)
    - 4\,\widehat{C}_{\rm mag}(m_Q)\,\Big[ (w-1)\,\chi_2^{\rm ren}(w)
    - 3 \chi_3^{\rm ren}(w) \Big] \,, \nonumber\\
   \phantom{ \Big[ } \xi_{\rm ren}(w)\,L_V(w) &=&
    2 \chi_1^{\rm ren}(w) - 4\,\widehat{C}_{\rm mag}(m_Q)\,
    \chi_3^{\rm ren}(w) \,, \nonumber\\
   \phantom{ \Big[ } \xi_{\rm ren}(w)\,L_3(w) &=&
    4\,\widehat{C}_{\rm mag}(m_c)\,\chi_2^{\rm ren}(w) \,.
\end{eqnarray}
$L_P$ and $L_V$ are corrections to the Isgur-Wise function which always
appear for pseudoscalar and vector mesons, respectively, irrespective
of the structure of the current \cite{FaNe}. We first present the
results for $N_+$ and $N_{A_1}$, which will play a special role in the
analysis below. They are:
\begin{eqnarray}\label{Rires1}
    N_+(w) &=& \bigg[ \widehat{C}_1(\bar w) + {w+1\over 2}\,
    \Big( \widehat{C}_2(\bar w) + \widehat{C}_3(\bar w) \Big)
    \bigg]\,\Big\{ 1 + \varepsilon_c L_D(w)
    + \varepsilon_b L_B(w) \Big\} \nonumber\\
   &&\mbox{} + \varepsilon_c\,{w-1\over 2}\,
    \Big\{ \big[ 1 - 2\eta(w) \big]\,\widehat{C}_2(\bar w) +
    \big[ 3 - 2\eta(w) \big]\,\widehat{C}_3(\bar w) \Big\}
    \nonumber\\
   &&\mbox{} + \varepsilon_b\,{w-1\over 2}\,
    \Big\{ \big[ 3 - 2\eta(w) \big]\,\widehat{C}_2(\bar w) +
    \big[ 1 - 2\eta(w) \big]\,\widehat{C}_3(\bar w) \Big\}
    \,, \nonumber\\
   && \\
   N_{A_1}(w) &=& \widehat{C}_1^5(\bar w)\,\Big\{ 1
    + \varepsilon_c L_{D^*}(w) + \varepsilon_b L_B(w) \Big\}
    \nonumber\\
   &&\mbox{} + \varepsilon_c\,{w-1\over w+1}\,\Big\{
    \widehat{C}_1^5(\bar w) + 2\eta(w)\,\widehat{C}_3^5(\bar w)
    \Big\} \nonumber\\
   &&\mbox{} + \varepsilon_b\,{w-1\over w+1}\,\Big\{
    \big[ 1 - 2\eta(w) \big]\,\widehat{C}_1^5(\bar w)
    + 2\eta(w)\,\widehat{C}_2^5(\bar w) \Big\} \,. \nonumber
\end{eqnarray}
Notice again that the Wilson coefficients are functions of the quark
velocity transfer $\bar w$, whereas the universal form factors depend
on the meson velocity transfer $w$. The expressions for the remaining
four form factors are more lengthy. To display them we omit the
dependence on $\bar w$ and $w$. We find:
\begin{eqnarray}\label{Rires2}
   N_- &=& {w+1\over 2}\,\big(\widehat{C}_2 - \widehat{C}_3\big)\,
    \Big\{ 1 + \varepsilon_c L_D + \varepsilon_b L_B \Big\} \nonumber\\
   &&\mbox{} - \varepsilon_c\,\bigg\{ (1-2\eta)\,\bigg[
    \widehat{C}_1 - {w-1\over 2}\,\big(\widehat{C}_2-\widehat{C}_3\big)
    \bigg] + (w+1)\,\widehat{C}_3 \bigg\} \nonumber\\
   &&\mbox{} + \varepsilon_b\,\bigg\{ (1-2\eta)\,\bigg[
    \widehat{C}_1 + {w-1\over 2}\,\big(\widehat{C}_2-\widehat{C}_3\big)
    \bigg] + (w+1)\,\widehat{C}_2 \bigg\} \,, \nonumber\\
   && \nonumber\\
   N_V &=& \widehat{C}_1\,\Big\{ 1 + \varepsilon_c L_{D^*}
    + \varepsilon_b L_B \Big\} \nonumber\\
   &&\mbox{} + \varepsilon_c\,\Big\{ \widehat{C}_1
    - 2\eta\,\widehat{C}_3 \Big\} + \varepsilon_b\,\Big\{
    \big[ 1 - 2\eta \big]\,\widehat{C}_1
    - 2\eta\,\widehat{C}_2 \Big\} \,, \nonumber\\
   && \\
   N_{A_2} &=& \widehat{C}_2^5\,\Big\{ 1 + \varepsilon_c L_{D^*}
    + \varepsilon_b L_B \Big\} + \Big[ \widehat{C}_1^5 + (w-1)\,
    \widehat{C}_2^5 \Big]\,\varepsilon_c L_3 \nonumber\\
   &&\mbox{} - {2\varepsilon_c\over w+1}\,\bigg\{
    (1+\eta)\,\big(\widehat{C}_1^5 + \widehat{C}_3^5\big)
    - {w+1\over 2}\,(1+2\eta)\,\widehat{C}_2^5 \bigg\} \nonumber\\
   &&\mbox{} + {2\varepsilon_b\over w+1}\,\widehat{C}_2^5\,
   \bigg\{ {3w+1\over 2} - (w+2)\,\eta \bigg\} \,, \nonumber\\
   && \nonumber\\
   N_{A_3} &=& \big(\widehat{C}_1^5 + \widehat{C}_3^5\big)\,
    \Big\{ 1 + \varepsilon_c L_{D^*} + \varepsilon_b L_B \Big\}
    - \Big[ \widehat{C}_1^5 - (w-1)\,\widehat{C}_3^5 \Big]\,
    \varepsilon_c L_3 \nonumber\\
   &&\mbox{} + {\varepsilon_c\over w+1}\,\Big\{ (w-1-2\eta)\,
    \big(\widehat{C}_1^5 - \widehat{C}_3^5\big)
    + 4w\,(1+\eta)\,\widehat{C}_3^5 \Big\} \nonumber\\
   &&\mbox{} + \varepsilon_b\,\bigg\{ (1-2\eta)\,
    \big(\widehat{C}_1^5 + \widehat{C}_3^5\big) + {2\over w+1}\,
    (w\eta-1)\,\widehat{C}_2^5 \bigg\} \,. \nonumber
\end{eqnarray}
These expressions are the main result of this paper. In
leading-logarithmic approximation, where $\widehat{C}_1=
\widehat{C}_1^5$ and all other coefficients are set to zero, they
reduce to the approximate expressions obtained by Luke~\cite{Luke}.

It is difficult to extract much information from the complicated
expressions for $N_i$ without a prediction for the subleading universal
form factors $\eta(w)$ and $\chi_i^{\rm ren}(w)$. Recently, these
functions have been investigated in great detail using the QCD sum rule
approach. The interested reader is referred to Refs.~\cite{xi3,sumrul}
for further discussion. Some important, general observations can be
made without such an analysis, however. Vector current conservation
implies that $L_P(1)=L_V(1)=0$, from which one can derive the
well-known normalization conditions $\xi_{\rm ren}(1)=1$ and
$\chi_1^{\rm ren}(1)= \chi_3^{\rm ren}(1)=0$. Inserting this into
(\ref{Rires1}), one finds that the $1/m_Q$ corrections in $N_+$ and
$N_{A_1}$ vanish at zero recoil. This is Luke's theorem \cite{Luke},
which implies that the leading power corrections to the meson form
factors $h_+(1)$ and $h_{A_1}(1)$ are of order $1/m_Q^2$. In
particular, it follows that
\begin{equation}
   h_{A_1}(1) = \widehat{C}_1^5(1) + {\cal{O}}(1/m_Q^2) \,.
\end{equation}
This relation plays a central role in the model-independent extraction
of $V_{cb}$ from $\bar B\to D^*\ell\,\bar\nu$ decays \cite{Vcb}. In
this case, even the second-order corrections have been analyzed in
detail and are found to be suppressed \cite{FaNe}.

It has been emphasized in Ref.~\cite{Riec} that Luke's theorem does not
protect the remaining form factors in (\ref{Rires2}). For instance, the
$\bar B\to D\,\ell\,\bar\nu$ decay amplitude at zero recoil is
proportional to the combination
\begin{equation}
   h_+(1) - \sqrt{S}\,h_-(1) = \Big[ \widehat{C}_1(1)
   + \widehat{C}_2(1) + \widehat{C}_3(1) \Big]
   \Big\{ 1 + S\cdot K \Big\} \,,
\end{equation}
where $S=\big({m_B-m_D\over m_B+m_D}\big)^2\approx 0.23$. The $1/m_Q$
corrections enter this expression through $h_-(1)$ and are contained
in the quantity $K$. Using the analytic expressions for the Wilson
coefficients given in Ref.~\cite{QCD}, we find
\begin{equation}
   K = \delta_1 + (\varepsilon_c + \varepsilon_b)\,
   \Big[ (1+\delta_1) - 2 (1+\delta_2)\,\eta(1) \Big] \,,
\end{equation}
with
\begin{eqnarray}
   \delta_1 &=& {1+z\over 1-z}\,\bigg\{ {4\alpha_s(m_c)\over 3\pi}
    - {2\alpha_s(m_b)\over 3\pi} + {4\alpha_s(\bar m)\over 3\pi}\,
    {z\over 1-z} \bigg( {\ln z\over 1-z} + 1 \bigg) \bigg\} \,,
    \nonumber\\
   && \\
   \delta_2 &=& {4\alpha_s(m_c)\over 3\pi}
    - {2\alpha_s(m_b)\over 3\pi} \,. \nonumber
\end{eqnarray}
Here $z=m_c/m_b$, and $\bar m$ denotes the geometric average of the
heavy quark masses ($\bar m\approx 2.23$ GeV). Note that $\delta_1$ has
a smooth limit as $z\to 1$. Numerically, we find $\delta_1\approx 6\%$
and $\delta_2\approx 9\%$, and thus
\begin{equation}\label{Kres}
   K\approx 0.3 - 0.5\,\eta(1) \,.
\end{equation}
Recently, the function $\eta(w)$ has been calculated using the QCD sum
rule approach, with the result that $\eta(w)=0.6\pm 0.2$ essentially
independent of $w$ \cite{xi3}. We observe that, by a fortunate
accident, this leads to an almost perfect cancellation in (\ref{Kres}),
i.e., $K\approx 0.0\pm 0.1$. This means that the $1/m_Q$ corrections to
the $\bar B\to D\,\ell\,\bar\nu$ decay rate at zero recoil are highly
suppressed.

Further predictions can be made for ratios of meson form factors, in
which some of the universal functions drop out \cite{Neu3}. An
important example is the ratio $R=h_V/h_{A_1}$, which can be extracted
from a measurement of the forward-backward asymmetry in $\bar B\to
D^*\ell\,\bar\nu$ decays. From (\ref{Rires1}) and (\ref{Rires2}) it is
readily seen that $R$ is independent of the functions $\chi_i^{\rm
ren}(w)$. In fact, at order $1/m_Q$ the following simple expression can
be derived:
\begin{equation}
   R = F_1(w)\,\bigg\{ 1 + {2\varepsilon_c\over w+1}
   + {2\varepsilon_b\over w+1}\,\Big[ 1 - 2 F_2(w)\,\eta(w) \Big]
   \bigg\} \,.
\end{equation}
Note that the form factor $\eta(w)$ enters in the $1/m_b$ corrections
only. The functions $F_i(w)$ contain the short-distance corrections and
are given by
\begin{eqnarray}
   F_1(w) &=& 1 + {4\alpha_s(m_c)\over 3\pi}\,r(w) \,, \nonumber\\
   F_2(w) &=& 1 + {2\alpha_s(\bar m)\over 3\pi}\,
    {(w^2-1)\,r(w) + (w-z)\,\ln z\over 1 - 2wz + z^2} \,,
\end{eqnarray}
where $r(w)=\ln(w+\sqrt{w^2-1})/\sqrt{w^2-1}$. The second function is
almost independent of $w$ over the kinematic range accessible in
semileptonic decays: $F_2(w)\approx 0.9$. Assuming that the sum rule
estimate $\eta(w)\approx 0.6$ is at least approximately correct, we
observe again a substantial cancellation: $1-2 F_2(w)\,\eta(w)\approx
-0.1$. This means that the $1/m_b$ corrections in $R$ can be safely
neglected, and up to terms of order $1/m_Q^2$ the form factor ratio can
be predicted in an essentially model independent way. Note that both
the QCD and the $1/m_c$ corrections are positive. As a consequence, the
deviations from the symmetry limit $R=1$ are rather substantial. For
instance, using $\bar\Lambda=0.5\pm 0.2$ GeV and neglecting the $1/m_b$
corrections we obtain $R=1.33\pm 0.08$ at zero recoil.

\section{Conclusions}

Starting from the observation that the structure of the short-distance
expansion of heavy quark currents is to a large extent determined by a
reparameterization invariance of HQET, we have derived the exact
expressions for the meson and baryon weak decay form factors to
next-to-leading order in $1/m_Q$. The results are presented in an
explicitly renormalization-group invariant form by introducing
renormalized Isgur-Wise form factors. The final formulas do not rely on
explicit expressions for the Wilson coefficient functions and are thus
valid to all orders in perturbation theory.

We have emphasized that beyond the leading order in $1/m_Q$ it is
necessary to distinguish between the hadron velocity transfer $w=v\cdot
v'$ and a related variable $\bar w$, which can be interpreted as the
short-distance velocity transfer of the heavy quarks. Whereas the
universal form factors of HQET are functions of $w$, the variable
$\bar w$ appears in the short-distance coefficient functions.

Our final expressions for the meson form factors in (\ref{Rires1}) and
(\ref{Rires2}), and for the baryon form factors in (\ref{Rires}),
generalize approximate results derived in leading-logarithmic
approximation by Luke \cite{Luke} and Georgi et al.~\cite{GGW}. These
new formulas should be used in further analyses of semileptonic decays
of heavy mesons or baryons.

\bigskip
{\it Acknowledgments:\/}
It is a pleasure to thank Michael Peskin for useful discussions.
Financial support from the BASF Aktiengesellschaft and from the German
National Scholarship Foundation is gratefully acknowledged. This work
was also supported by the Department of Energy, contract
DE-AC03-76SF00515.

\end{document}